\documentclass{emulateapj}

\usepackage{xspace}
\usepackage{graphicx}
\usepackage{natbib}
\usepackage{apjfonts}
\def\errtwo#1#2#3{$#1^{+#2}_{-#3}$}

\newcommand\fu{{4U~1957+11}\xspace}

\newcommand\chandra{\textsl{Chandra}\xspace}

\newcommand\gso{{GSO}\xspace}

\newcommand\hetg{{HETG}\xspace}

\newcommand\hxd{{HXD}\xspace}
\newcommand\isis{{\tt ISIS}\xspace}

\newcommand\pin{{PIN}\xspace}

\newcommand\rxte{\textsl{RXTE}\xspace}

\newcommand\suzaku{\textsl{Suzaku}\xspace}

\newcommand\xis{{XIS}\xspace}
\newcommand\xmm{\textsl{XMM-Newton}\xspace}

\newcommand\aproxgt{\mathrel{%
     \rlap{\raise 0.511ex \hbox{$>$}}{\lower 0.511ex \hbox{$\sim$}}}}
\newcommand\aproxlt{\mathrel{%
     \rlap{\raise 0.511ex \hbox{$<$}}{\lower 0.511ex \hbox{$\sim$}}}}

\slugcomment{Submitted 2011 May 1}

\shorttitle{Suzaku Observations of \fu}
\shortauthors{Nowak et al.}

\begin{document}

\title{Suzaku Observations of \fu: Potentially the Most Rapidly Spinning
Black Hole \\ in (the Halo of) the Galaxy}

\author{Michael A. Nowak\altaffilmark{1}, 
  J\"orn Wilms\altaffilmark{2}, Katja
  Pottschmidt\altaffilmark{3}, Norbert Schulz\altaffilmark{1}, Dipankar
  Maitra\altaffilmark{4}, Jon Miller\altaffilmark{4}} 
\altaffiltext{1}{Massachusetts Institute of
  Technology, Kavli Institute for Astrophysics, Cambridge, MA 02139,
  USA; mnowak,nss@space.mit.edu}
\altaffiltext{2}{Dr.~Karl Remeis-Sternwarte and Erlangen Centre for
  Astroparticle Physics, Universit\"at Erlangen-N\"urnberg,
  Sternwartstr.~7, 96049 Bamberg, Germany; 
  joern.wilms@sternwarte.uni-erlangen.de}
\altaffiltext{3}{CRESST, UMBC, and
  NASA Goddard Space Flight Center, Greenbelt, MD 20771;
  katja@milkyway.gsfc.nasa.gov} 
\altaffiltext{4}{Dept. of Astronomy, Univ. of Michigan, 
  500 Church St., Ann Arbor, MI  48109-1042; dmaitra,jonmm@umich.edu}

\begin{abstract}
We present three \suzaku observations of the black hole candidate \fu
(V1408~Aql) --- a source that exhibits some of the simplest and
cleanest examples of soft, disk-dominated spectra.  \fu also presents
among the highest peak temperatures found from disk-dominated
spectra. Such temperatures may be associated with rapid black hole
spin.  The \fu spectra also require a very low normalization, which
can be explained by a combination of small inner disk radius and a
large distance ($>10$\,kpc) which places \fu well into the Galactic
halo.  We perform joint fits to the \suzaku spectra with both
relativistic and Comptonized disk models.  Assuming a low mass black
hole and the nearest distance (3\,$M_\odot$, 10\,kpc), the
dimensionless spin parameter $a^* \equiv Jc/GM^2 \gtrsim 0.9$.  Higher
masses and farther distances yield $a^* \approx 1$.  Similar
conclusions are reached with Comptonization models; they imply a
combination of small inner disk radii (or, equivalently, rapid spin)
and large distance.  Low spin cannot be recovered unless \fu is a low
mass black hole that is at the unusually large distance of $\gtrsim
40$\,kpc.  We speculate whether the suggested maximal spin is related
to how the system came to reside in the halo.
\end{abstract}


\section{Introduction}\label{sec:intro}

\setcounter{footnote}{0}

\fu is one of the few black hole candidates (BHC), like LMC~X-1,
LMC~X-3, and Cyg~X-1, that historically has been persistently
active. In contrast to these other sources, however, it is a Low Mass
X-ray Binary \citep[LMXB;
  see][]{thorstensen:87a,hakala:99a,russell:10a,bayless:10a}.  Also in
contrast to other persistent Galactic BHC sources, \fu apparently has
remained always in a spectrally soft state \citep{nowak:08a}.  Its
soft spectrum is well-modeled as a simple disk
\citep{nowak:99d,wijnands:02c,nowak:08a,dunn:10a}, i.e., a
multi-temperature blackbody \citep{mitsuda:84a} characterized by a
peak temperature and a normalization related to the disk inner radius,
inclination, and object distance.  Approximately 15\% of the over 40
pointed observations taken with the {\textsl Rossi X-ray Timing
  Explorer} (\rxte), however, also show a significant contribution
(20--75\% of the observed 3--18\,keV flux) from a non-thermal/steep
powerlaw component \citep{nowak:08a,dunn:10a}.

Optical observations suggest that we may be viewing the disk in \fu at
a high inclination of $\sim 75^\circ$. Modulation over a 9.33\,hr
orbital period has ranged from $\pm 10\%$ and sinusoidal
\citep{thorstensen:87a} to $\pm30\%$ and complex
\citep{hakala:99a}. The latter has been interpreted as a high
inclination warped disk being partly occulted by the secondary.  On
the other hand, \citet{russell:10a} failed to find any optical
modulation over the orbital period, while \citet{bayless:10a} found
nearly sinusoidal modulation.  { These latter authors attributed
  the optical modulation to illumination of the secondary, and noting a lack
  of any flattening at the lightcurve minimum, they modeled the system
  inclination with values} as low as $20^\circ$ and not greater than
$70^\circ$.

Currently, the mass, distance, and inclination of \fu are unknown.
Based upon a comparison of the optical flux to the estimated optical
luminosities of other BHC, however, \citet{russell:10a} have argued
that \fu must lie at a distance $>20$\,kpc.  { The possibility of a
  high inclination, however, makes such comparisons
  problematic. \citet{russell:11a} recently have presented
  near-simultaneous radio (\textsl{Expanded Very Large Array} --
  \textsl{EVLA}), optical (\textsl{Faulkes Telescope North}), and
  UV/X-ray (\textsl{Swift}) observations of \fu.  These observations
  show the soft state flux to be extremely quenched (factors of
  330--810 times lower than the expected radio/X-ray flux ratio for
  hard state black hole sources), and demonstrate that the optical/UV
  flux are roughly consistent with the expectation of an irradiated
  disk.  Those observations, however, do not definitively constrain
  the source's inclination or distance.}  A minimum distance of
$>5$\,kpc; { however,} has been determined via high resolution
X-ray spectroscopy.  Ne\,{\sc ix} $13.45$\,\AA\ absorption, associated
with the warm/hot phase of the interstellar medium, is detected with a
sufficient equivalent width to suggest that \fu resides above the
galactic plane at this minimum distance or beyond
\citep{yao:08a,nowak:08a}.

\citet{nowak:08a} showed that, with the exception of the $\approx15$\%
of the time when a non-thermal component becomes prominent, disk
models of the soft X-ray component have an approximately constant
radius but a temperature that increases with flux (see Fig.~9 of
\citealt{nowak:08a}).  There is some indication for a slight increase
in disk radius at the lowest flux levels, which was interpreted as
possibly being associated with the source approaching a transition to
a spectrally hard state.  Such a transition is expected to occur at
2--3\%\,$\mathrm{L_{Edd}}$, where $\mathrm{L_{Edd}}$ is the Eddington
luminosity for the source in question \citep{maccarone:03a}. At the
other extremes of flux in \fu, the simple disk spectrum ``drops out''
(disk radius decreases in disk+powerlaw fits, or disk temperature
decreases in disk+Compton corona fits) and is replaced by the
non-thermal/steep powerlaw component \citep{nowak:08a}.

Although other persistent systems transiently show disk-dominated
spectra\footnote{LMC~X-1 is likely wind-fed, and, for a soft state, 
  shows an unusual and highly variable disk spectrum
  \citep{nowak:01a}. LMC~X-3 cycles through soft/hard state
  transitions \citep{wilms:01a}.}, the persistent, soft spectra of \fu
are perhaps the simplest and cleanest examples of ``disk spectra''.
\xmm and \chandra observations show that there is very little
absorption ($N_\mathrm{H} = 1$--$2\times10^{21}~\mathrm{cm^{-2}}$),
while \rxte observations show that a hard tail contributes usually
$<20\%$ of the flux. Furthermore, any additional components such as a
broad Fe line (typically $<2\%$ peak residual in the fits) or a
``smeared Fe edge'' (which was never required or indicated in any of
the \rxte spectra; \citealt{nowak:08a}) are very weak or absent.  Thus
the soft spectrum of \fu becomes the ideal testbed for modern disk
atmosphere models that incorporate spin and other General Relativistic
effects into their calculations
\citep[e.g.,][]{li:05a,davis:05a,davis:06a,shafee:06a}.

\citet{nowak:08a} showed that the disk-fits to the \xmm, \chandra, and
\rxte spectra of \fu are characterized by a very low normalization,
indicating some combination of large distance and low compact object
mass, and very high inner disk temperature (1.3--1.8\,keV).  In other
sources the latter has been associated with high black hole spin.  To
assess this possibility, \citet{nowak:08a} applied the relativistic
disk model, {\tt kerrbb}, of \citet{li:05a}.  The model parameters
include a compact object mass and distance, disk inclination,
accretion rate, spectral hardening factor (the ratio of color
temperature to effective temperature), and dimensionless spin of the
black hole, $a^*\equiv Jc/GM^2$.  For a wide variety of masses and
distances, fits to the \xmm and \chandra spectra preferred maximal
spin, $a^* \approx 1$.  The \rxte spectra, in fact, could not be
described with low spin, even when including a coronal or powerlaw
component.

Various theoretical models of a disk atmosphere have suggested a
spectral hardening factor of $\sim1.7$.  If this is correct, and
utilizing the fact that the lowest flux observations must be $\aproxgt
3\%\,\mathrm{L_{Edd}}$, then the spectral fits of \fu strongly suggest
that it is a maximally spinning, $\approx$16\,$M_\odot$ black hole at
a distance of $\approx$22\,kpc \citep{nowak:08a}. In this work we
explore that claim with three recent \suzaku observations.

In Section~\ref{sec:data} we describe the \suzaku observations and the
details of the data processing.  We then present the model fits to the
resulting spectra in Section~\ref{sec:fits}.  Section~\ref{sec:scale}
describes the scaling relations that we used to translate fits using a
low mass, short distance black hole to comparably good fits involving
a larger mass, further distance black hole.  We then present our
conclusions in Section~\ref{sec:discuss}.

\begin{deluxetable}{cccc}
\setlength{\tabcolsep}{0.03in} 
\tabletypesize{\footnotesize}    
\tablewidth{0pt} 
\tablecaption{Log of \fu Observations \label{tab:data}}
\tablehead{\colhead{Date} &
          & \colhead{ObsID}
          & {Exposure}
          \\                               
          (yyyy-mm-dd) & (MJD) & & {(ksec)}
         }
\startdata
    2010-05-04 & 55320.82 & 405057010 &  35.80
\\

    2010-05-17 & 55333.80 & 405057020 &  37.20
\\

    2010-11-01 & 55502.19 & 405057030 & 15.50 \enddata
    \tablecomments{Exposure times are after good time filtering, and
      represent the maximum exposure in an \xis detector when
      combining all data modes. { MJD is the (non-barycenter
        corrected) average Modified Julian Date of the observation.}}
\end{deluxetable} 

\section{Observations and Data Analysis}\label{sec:data}

The \suzaku data were reduced with tools from the HEASOFT v6.9 package
and the calibration files current as of 2010 September. The
instruments on \suzaku \citep{mitsuda:07a} are the X-ray Imaging
Spectrometer \citep[\xis;][]{koyama:07a} CCD detector covering the
$\approx 0.3$--10\,keV band, and the Hard X-ray Detector
\citep[\hxd;][]{takahashi:07a} comprised of the PIN diode detector
(\pin) covering the $\approx 10$--70\,keV band and the gadolinium
silicate crystal detector (\gso) covering the $\approx~60$--600\,keV
band. The \xis had four separate detectors, \xis0--3, with \xis1 being
a backside illuminated CCD. \xis2 was lost due to a micrometeor hit in
late 2006, and thus we only consider data from the remaining three
\xis detectors.  For all three observations described here, the hard
X-ray fluxes were too low to lead to a significant detection in either
of the \hxd detectors, and therefore we do not include these data. A
log of the observations is found in Table~\ref{tab:data}.

\begin{figure}
\epsscale{1} \plotone{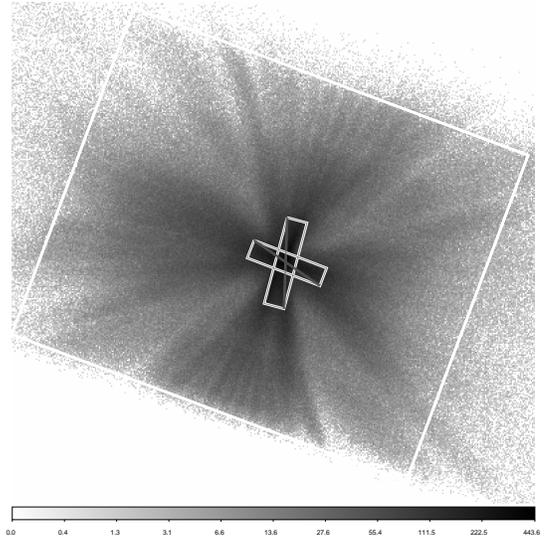}
\caption{The { aspect corrected} \xis1 { 0.5--8\,kev} image on a
  logarithmic intensity scale from ObsID 405057010.  The boxes outline
  the detector region from which the spectra were generated.  The
  length of the short edge of the outer box is approximately
  $4^\prime.5$. The inner boxed regions with diagonal slashes were
  excluded { from the spectra}.}
 \label{fig:ds9}
\end{figure}

All of the \xis observations were run in a sub-array mode wherein 1/4
of the CCD was exposed with 2 second integration times.  In preparing
the \xis spectra, we first corrected each detector for Charge Transfer
Inefficiency (CTI) using the \texttt{xispi} tool, and then reprocessed
the data with \texttt{xselect} using the standard \texttt{xisrepro}
selection criteria. Due to thermal flexing of the spacecraft, the
attitude of the \suzaku spacecraft exhibits variability over the
course of the observations and therefore the image of the source is
not at a fixed position on the CCD. Standard processing reduces this
variability and improves the Point Spread Function (PSF) image
\citep{uchiyama:08a}; however, the standard tools do not yet fully
correct the variable attitude induced blurring.  We created an
improved attitude reconstruction, and hence narrower PSF images, using
the \texttt{aeattcor.sl} software described\footnote{See also: {\tt
    http://space.mit.edu/CXC/software/suzaku/} {\tt aeatt.html}.} by
\citet{nowak:10a}.  For the third observation, however, approximately
half of the good time intervals (i.e., half of each spacecraft orbit)
were affected by wobbles severe enough ($>1^\prime$) such that a large
fraction of the image on the \xis1 detector fell beyond the boundaries
of the detector's active regions.  Although the images did not fall
beyond the active region boundaries on the \xis0 and \xis3 detectors,
there are serious questions about the accuracy of the generated
response functions for such large off axis angles.  We therefore
excluded these times from the spectra, and reflect this in
Table~\ref{tab:data}.

\begin{figure}
\epsscale{1} \plotone{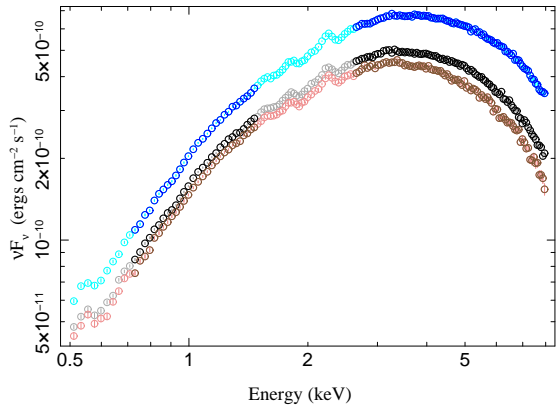}
\caption{The combined \xis0, 1, and 3 spectra from each of the three
  \suzaku observations of \fu, flux corrected with the detector
  responses.  (ObsID 405057010 is the top spectrum, obsID 405057020 is
  the middle spectrum, and obsID 405057030 is the bottom spectrum.)
  The three observations share a common binning of approximately half
  width half maximum resolution of the detectors. { Lighter colored
    data points represent energy bands that were ignored in the
    spectral fits described below.}}
 \label{fig:data}
\end{figure}

Using the image generated from the improved attitude correction, we
then used the \texttt{pile\_estimate.sl} \texttt{S-Lang}
script\footnote{{\tt
    http://space.mit.edu/CXC/software/suzaku/pest.html}}
\citep{nowak:10a} to estimate the degree of pileup in the \xis
observations. For the spectra described in this work, the center of
the PSF images were affected by as much as a 33\% pileup fraction.  We
therefore exclude two, overlapping rectangular regions ($\approx
15\times60$ pixels, nearly at right angles to one another) aligned
along the brightest portions of the central PSF (see
Fig.~\ref{fig:ds9}).  We estimate that the remaining residual pileup
fractions are $\lesssim 3\%$.

The outer limit of the extraction area was a rectangular region
limited by the $4^\prime$ width of the 1/4 array boundaries.
Additionally, we extracted background spectra from small rectangular
regions, near the edges of the chips, that avoided as much as possible
both the \fu image as well as the detector images of the on-board
calibration sources.  It is not entirely clear to what degree these
regions are dominated by ``true background'', as opposed to the wings
of the PSF from the source.  However, the normalized background flux
remains at $<2\%$ of the source flux between 0.5--8\,keV, and thus
does not strongly affect the analyses presented in this work.  The
background flux starts to become an increasingly larger fraction of
the source flux above 8\,keV.  For this reason, as well as due to any
other uncertainties in the high energy calibration of the \xis
detectors, we do not consider data above 8\,keV.

Events in the \xis detectors were read out from either $3\times3$ or
$5\times5$ pixel islands.  We created individual spectra and response
files for each detector and data mode combination.  Little source
variability was detected over the course of each observation,
therefore the spectra were created for the full good time intervals
represented by standard processing (excluding the times of very bad
attitude errors, discussed above). Response matrices and effective
area files were created with the \texttt{xisrmfgen} and
\texttt{xissimarfgen} tools, respectively.  For each observation,
spectra for the different readout modes were combined during analysis
on a chip-by-chip basis using the {\tt combine\_datasets} function
within the {\tt Interactive Spectral Interpretation System} (\isis)
fitting program \citep{houck:00a}.  That is, observations were kept
separate, but within an observation all \xis0 spectra were combined
together, all \xis1 spectra were combined, etc.

Although spectra for each observation and each \xis chip were kept
separate, all spectra were jointly grouped on a common grid such that
the \xis0 spectra for the third (i.e., the faintest and shortest
integration time) observation had a minimum combined signal-to-noise
ratio of 5 in each energy bin ($\approx 25$ total counts in each bin,
although the estimated background was included in the S/N criterion)
and that the minimum number of channels per energy bin was at least
the half width half maximum of the spectral resolution \citep[see][for
  a further explanation of this criterion]{nowak:10a}.

In Fig.~\ref{fig:data} we present the flux corrected spectra, i.e.,
the counts spectra divided by the integrated response (see the \isis
user manual for a full description of this process).  The spectra show
strong deviations from a smooth continuum in the 1.5--2.6\,keV region,
near X-ray spectral features of Si and Ir.  These systematic effects
are seen in a wide variety of unrelated \suzaku spectra, hence our
decision to exclude them from our fits.  Likewise, there is a strong
spectral feature near 0.55\,keV, which we also suspect of being
systematic in nature, hence our exclusion of data below 0.7\,keV. To
avoid these regions of poorly understood response, we only considered
spectral energy ranges 0.7--1.5\,keV and 2.6--8\,keV.  The smaller
features near 0.75\,keV and 0.9\,keV are plausibly associated with Fe
L, Ne K, and Ne {\sc ix} absorption in the interstellar medium (ISM),
as we discuss below.

\begin{figure}
\epsscale{1} \plotone{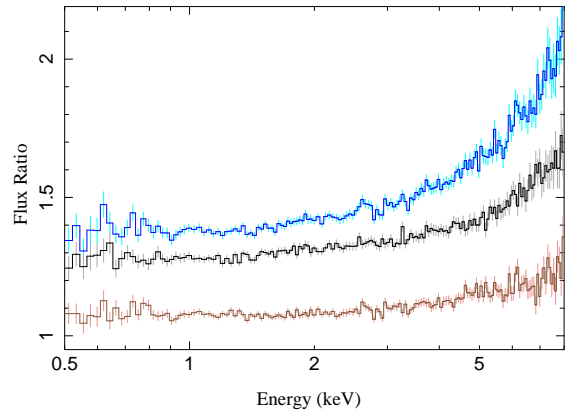}
\caption{Ratios of the flux corrected spectra shown in
  Fig.~\protect{\ref{fig:data}}.  The top ratio is observation
  1/observation 3, the middle line is observation 1/observation2, and
  the bottom line is observation 2/observation 3.}
 \label{fig:ratio}
\end{figure}

\begin{figure}
\epsscale{1} \plotone{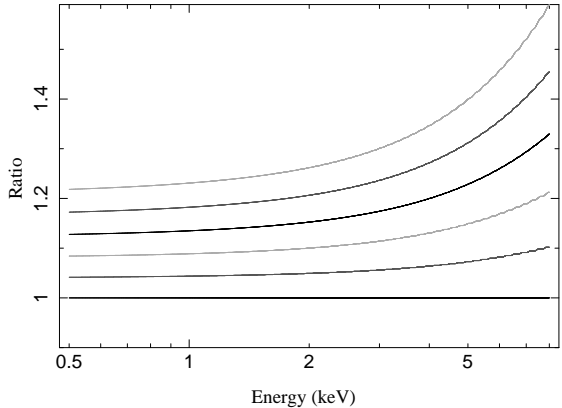}
\caption{Expectations for the flux ratio in a {\tt diskpn} model when
  solely changing the peak temperature of the disk.  The reference
  model has a 1.3\,keV peak temperature.  The peak temperature is
  varied in steps of 0.05\,keV, up to 1.5\,keV.}
 \label{fig:tratio}
\end{figure}

The suspected systematic features are relatively constant among the three
observations, as is evident from the ratio of the flux corrected
spectra as shown in Fig.~\ref{fig:ratio}.  None of the systematic
features are seen in these ratios.  The ratios are relatively flat
$\lesssim 2$\,keV, and show a curve upward, with the degree of this
curving being greatest for the ratio of the flux corrected spectra for
the first and third observation.  We discuss this further in the next
section.

\begin{figure}
\epsscale{1} \plotone{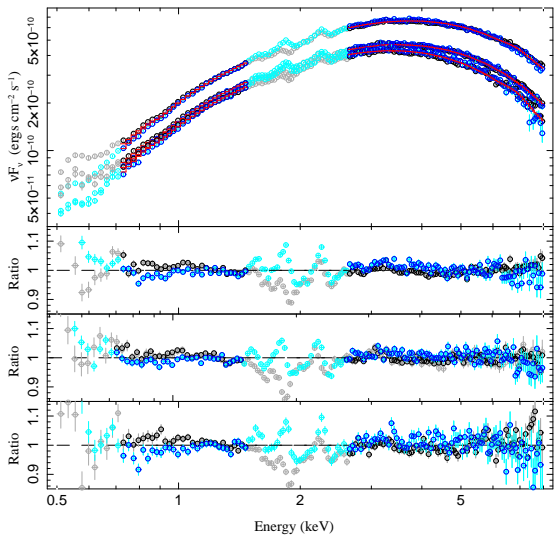}
\caption{The joint fit of an {\tt eqpair} Comptonization model to the
  three spectra.  The spectra share a common interstellar absorption
  column and a common normalization (i.e., disk inner radius), but
  have independent disk temperatures and coronal
  parameters. Black/grey spectra are the combined \xis0 and 3 spectra,
  and blue/light blue are the \xis1 spectra.  Lightly shaded data were
  not included in the fit.  The top residuals panel shows the ratio
  residuals for obsID 405057010, the middle residuals panel shows the
  ratio residuals for obsID 405057020, and the bottom panel shows the
  ratio residuals for obsID 405057030.}
 \label{fig:eqpair}
\end{figure}

\section{Spectral Fits}\label{sec:fits}

The spectral shape shown in Fig.~\ref{fig:data}, and the ratios seen
in Fig.~\ref{fig:ratio}, are those expected for a disk-dominated
spectrum wherein the flux changes are primarily dominated by
variations of the peak temperature of the disk.  We show the
theoretical expectation for such a disk in Fig.~\ref{fig:tratio}.
Here we use the {\tt diskpn} model \citep{gierlinski:99a}, which
essentially is the spectrum from a Shakura-Sunyaev type disk
\citep{shakura:73a} calculated using a simple pseudo-Netwonian
potential and a no-torque boundary condition at the disk inner edge
located at $6\,GM/c^2$.  The flux goes to zero at the disk inner edge,
and the disk temperature peaks at $\approx 10\,GMc^2$.  This basic
disk spectrum will underly the Comptonization fits discussed below.

\begin{figure}
\epsscale{1} \plotone{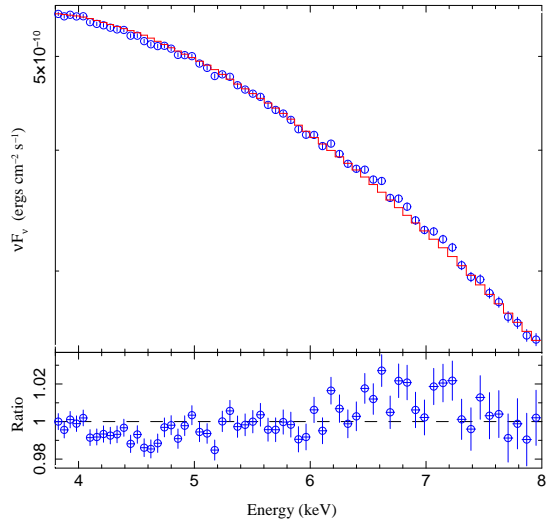}
\caption{A simultaneous fit to all three observations using the {\tt
    eqpair} model, but excluding the broad Gaussian line in the Fe
  region.  (The spectra were combined for the plot, but not the
  fits.}
 \label{fig:line}
\end{figure}

At energies approximately below the peak temperature, dominated by the
Rayleigh-Jeans part of the thermal spectra, the ratios are fairly
flat.  At energies above the disk peak temperature, on the Wien tail
of the thermal spectra, the ratios curve upward.  There is a very good
correspondence between the curves in Figs.~\ref{fig:ratio} and
\ref{fig:tratio}.  However, based upon our experience with \rxte
spectra which extended out to $\approx 20$\,keV, there is the
possibility of additional spectral hardening beyond that associated
with any rise in disk temperature.  We therefore consider the {\tt
  eqpair} Comptonization model \citep{coppi:99a} which can use the
{\tt diskpn} as its input seed photon spectrum.

\begin{deluxetable*}{lccccccccccccc}
\setlength{\tabcolsep}{0.03in} 
\tabletypesize{\footnotesize}    
\tablewidth{0pt} 
\tablecaption{{\tt constant*TBnew*(eqpair+gaussian)} Fits to \fu 
Spectra \label{tab:efits}}
\tablehead{
     \colhead{ObsID}
   & \colhead{$N_\mathrm{H}$}
   & \colhead{$N_\mathrm{eqp}$} 
   & \colhead{$kT_\mathrm{disk}$} 
   & \colhead{$\ell_\mathrm{h}/\ell_\mathrm{s}$} & \colhead{$\tau_\mathrm{net}$}
   & \colhead{$N_\gamma$} 
   & \colhead{$C_\mathrm{xis0}$} & \colhead{$C_\mathrm{xis3}$}
   & \colhead{0.7--8\,keV Flux}
   & \colhead{$\chi^2/$DoF}
          \\                               
   & ($10^{21}~\mathrm{cm^{-2}}$) & ($10^{-4}$) & (keV) 
   & & & ($10^{-4}$) & & & ($10^{-9}\,\mathrm{erg\,cm^{-2}\,s}$)
         }
\startdata
   405057010
 & \errtwo{0.200}{0.001}{0.001}     
 & \errtwo{1.926}{0.001}{0.001}     
 & \errtwo{1.477}{0.001}{0.001}     
 & \errtwo{0.37}{0.03}{0.02}        
 & \errtwo{0.017}{0.002}{0.002}     
 & \errtwo{1.4}{0.4}{0.5}           
 & \errtwo{1.019}{0.002}{0.002}     
 & \errtwo{1.020}{0.002}{0.002}     
 & $1.09\pm0.01$
 & 1750.4/1014
\\
\noalign{\vspace*{0.2mm}}
   405057020
 & \nodata
 & \nodata
 & \errtwo{1.342}{0.001}{0.001}     
 & \errtwo{0.05}{0.01}{0.01}        
 & \errtwo{2.4}{0.1}{0.1}        
 & \errtwo{0.1}{0.2}{0.1}           
 & \errtwo{1.078}{0.002}{0.002}     
 & \errtwo{1.083}{0.002}{0.002}     
 & $0.80\pm0.04$
 & \nodata
\\
\noalign{\vspace*{0.2mm}}
   405057030
 & \nodata
 & \nodata
 & \errtwo{1.310}{0.001}{0.001}     
 & \errtwo{0.04}{0.01}{0.01}        
 & \errtwo{4.4}{0.1}{0.1}        
 & \errtwo{1.3}{0.4}{0.4}           
 & \errtwo{1.067}{0.002}{0.002}     
 & \errtwo{1.057}{0.003}{0.003}     
 & $0.72\pm0.03$
 & \nodata
\\
\noalign{\vspace*{3mm}}
   405057010
 & \errtwo{0.199}{0.007}{0.008}     
 & \errtwo{1.926}{0.002}{0.002}     
 & \errtwo{1.476}{0.001}{0.001}     
 & \errtwo{0.38}{0.03}{0.03}        
 & \errtwo{0.017}{0.002}{0.002}     
 & \errtwo{1.4}{0.4}{0.4}           
 & \errtwo{1.020}{0.002}{0.002}     
 & \errtwo{1.020}{0.003}{0.002}     
 & \nodata
 & 1016.9/1014
\\
\noalign{\vspace*{0.2mm}}
   405057020
 & \nodata
 & \nodata
 & \errtwo{1.343}{0.001}{0.001}     
 & \errtwo{0.05}{0.01}{0.01}        
 & \errtwo{2.4}{0.1}{0.1}        
 & \errtwo{0.1}{0.3}{0.1}           
 & \errtwo{1.077}{0.002}{0.002}     
 & \errtwo{1.081}{0.003}{0.001}     
 & \nodata
 & \nodata
\\
\noalign{\vspace*{0.2mm}}
   405057030
 & \nodata
 & \nodata
 & \errtwo{1.310}{0.001}{0.001}     
 & \errtwo{0.05}{0.01}{0.01}        
 & \errtwo{4.5}{0.1}{0.1}             
 & \errtwo{1.3}{0.4}{0.4}             
 & \errtwo{1.066}{0.003}{0.004}     
 & \errtwo{1.056}{0.004}{0.004}     
 & \nodata
 & \nodata
\enddata 
\tablecomments{Errors are 90\% confidence level for one interesting
  parameter, except for the 0.8--8\,keV (absorbed) flux error which is
  from the standard deviation of the \xis detector cross
  normalizations.  $N_\mathrm{H}$ is the neutral column;
  $N_\mathrm{eqp}$ is the {\tt eqpair} normalization (see text);
  $kT_\mathrm{disk}$ is the peak temperature of the disk seed photons;
  $\ell_\mathrm{h}/\ell_\mathrm{s}$ is the relative corona to disk
  compactness ($\ell_\mathrm{s}$ is fixed to 1); $\tau_\mathrm{net}$
  is the net coronal optical depth (including pair production);
  $N_{\gamma}$ is the Gaussian line normalization in units of
  photons\,$\mathrm{cm^{-2}}~\mathrm{s}^{-1}$; and $C_\mathrm{xisN}$
  are the normalization constants for the \xis0 and 3 detectors,
  relative to the \xis1 detector. { The first group of parameters
    are for data that include only statistical errors, while the
    second group of parameters are for data that also include 1.35\%
    systematic errors.}}
\end{deluxetable*}  

Given that the three spectra appear commensurate with a constant disk
radius, and that such a nearly constant disk radius is consistent with
our experience for $\approx85$\% of the \rxte spectra that we have
previously examined (see Fig.~9 of \citealt{nowak:08a}), we explore a
model wherein we fit all three spectra simultaneously with an {\tt
  eqpair} model for which we constrain the {\tt eqpair} normalization
(i.e., the inner disk radius) to be the same for each spectrum.  {
  The physical implications of this common normalization are discussed
  in \S\ref{sec:discuss}.}

{ In addition to constraining the {\tt eqpair} normalization, we
  also} constrain the absorption to be the same for each observation,
which we here model with the {\tt TBnew} model (an updated version of
the model of \citealt{wilms:00a}, using the interstellar abundances
from that work) and using a fixed Ne {\sc ix} 13.447\,\AA\ absorption
line with equivalent width of 7\,\AA.  The latter is consistent with
the high spectral resolution studies of \citet{yao:08a} and
\citet{nowak:08a} (see the below discussion of the ISM absorption).
Excluding this component increases the fitted $\chi^2$ by $\approx
12$.

For each observation we allow the disk peak temperature,
$kT_\mathrm{disk}$, the coronal compactness,
$\ell_\mathrm{h}/\ell_\mathrm{s}$, and the coronal seed optical
depth\footnote{Table~\ref{tab:efits}, however, presents the \emph{net}
  optical depth, as {\tt eqpair} includes calculation of the
  equilibrium pair production in the corona.}, $\tau_s$, to be
independent parameters.  We include Gaussian lines to the fits;
however, as their presence is not strongly confirmed and there is no
good evidence for a narrow line in either these spectra nor in the
previously examined \chandra-\hetg spectra \citep{nowak:08a}, we tie
all the line energies (constrained to lie between 6.2--6.9\,keV) and
widths (fixed to 0.3\,keV), and let { only} the line normalizations
remain free.

The results of these fits are presented in Fig.~\ref{fig:eqpair}, and
the parameters are given in Table~\ref{tab:efits}.  Overall, the
quality of the fits is very good, yielding $\chi^2_\nu/\nu =
1756/1014=1.74$.  Although the reduced $\chi^2$ is not 1, the addition
of only 1.35\% systematic uncertainties to each dataset { brings
  the value down to that level, without significantly altering the
  best fit parameters, as also shown in Table~\ref{tab:efits}.  The
  addition of systematic uncertainties slightly increases the
  parameter error bars.}

Given the obvious systematic differences between the frontside (\xis0
and 3) and backside (\xis1) detectors, seen in the residuals presented
in Fig.~\ref{fig:eqpair}, we speculate that { our fits are near
  ``optimal'.  Throughout the rest of this work, the figures will
  include only statistical errors; however, tables will include sets
  of parameters with error bars that were calculated for spectra to
  which 1.35\% systematic errors have been applied.  This systematic
  error amplitude was chosen on the basis that it yields a reduced
  $\chi^2$ value of $\approx 1$ for the fits described above, and we
  retain this level in subsequent model fits for ease of comparison.}

The primary difference among the three spectra are the disk peak
temperatures, which range between 1.31--1.48\,keV.  The coronal
contribution is relatively weak, given the low values of the coronal
compactness compared to the disk compactness,
$\ell_\mathrm{h}/\ell_\mathrm{s}$.  It is interesting to note that the
two fainter observations indicate less energetic coronae ---
$\ell_\mathrm{h}/\ell_\mathrm{s}$ values are lower --- but higher
optical depth ones.  { (Coronal compactness is its input energy
  divided by a characteristic radius, here measured relative to a
  compactness for the disk.  If geometric changes are absent, then a
  lower fitted compactness is indicating a less energetic corona.)}
The resulting equilibrium coronal temperatures for the three
observations are $\approx 600$\,keV, $\approx 2.2$\,keV, and $\approx
1.3$\,keV, respectively.  When combined with their equilibrium optical
depths, we find that the resulting Compton $y$-parameters range from
0.1--0.2.  That is, Comptonization is only inducing a modest shift in
the average photon energy. However, given the fact that we do not have
a good spectrum above 8\,keV, there is not a strong enough lever arm
on these Comptonization fits to say how free from systematic
uncertainty these parameter values are.

Likewise, detections of a (broad) Gaussian line peaking between
6.2--6.9\,keV are marginal.  The upper limits of the line strengths
found here (in terms of integrated photon flux) are comparable to the
estimated values from our prior \rxte observations, after taking into
account the fact that \rxte also integrates line emission from the
galactic ridge and likely had remaining systematic Fe line residuals
at the $\approx 1\%$ level \citep{nowak:08a}.  That is, these line
strengths are marginally consistent with the \rxte results.
Furthermore, the line fitted here may be partly substituting for
general spectral hardening at $\gtrsim6$\,keV.  It is clear, however,
that any real line at the levels suggested by the \rxte observations
cannot be narrow; any such line must be broad and have a peak residual
relative to the continuum of $\lesssim 2\%$.  We illustrate this in
Fig.~\ref{fig:line}, wherein we have refit the spectra excluding the
Fe line, and combine all three observations in the plot --- but not
the fit --- to show the Fe region residuals.

The most salient feature of these fits is that a model with a single
normalization constant is both successful, and yields a fairly small
normalization of $\approx 1.9\times10^{-4}$.  The {\tt eqpair} model
normalization is proportional to the disk flux that would be observed
in the absence of Comptonization, and is { the same as the
  normalization of the {\tt diskpn} model \citep{gierlinski:99a},
  namely}
\begin{equation}
   N_\mathrm{eqp} = \left ( \frac{M}{1\,M_\odot} \right )^2 \left(
   \frac{D}{1\,\mathrm{kpc}} \right )^{-2} f_\mathrm{c}^{-4}\cos i ~~,
\label{eq:scale}
\end{equation} 
where $M$ is the central object mass, $D$ is the distance, $i$ is the
disk inclination, and $f_\mathrm{c}$ is the color-correction factor of
the disk { \citep[see also][]{coppi:99a}}.  For a mass of
$3\,M_\odot$, a distance of 10\,kpc, and an inclination of
$i=75^\circ$, we expect a normalization of $\approx 3\times10^{-3}$, a
factor of 15 times higher than observed.  (The discrepancy only
increases for a more face on geometry; a more highly inclined geometry
is unlikely given the lack of eclipses.) The fitted normalization can
be recovered if we increase the distance to 40\,kpc, increase the
color-correction factor for the seed input photons to 3.3, or we
decrease the characteristic emission radius by a factor of four.  (The
emitting area dependence is responsible for the $M^2$ term in the
normalization.)

The {\tt diskpn} model, which provides the seed photons for the {\tt
  eqpair} model,  has its peak emission radius at $\approx 10\,GM/c^2$.
If this radius were shrunk to $2.5\,GM/c^2$, the fitted normalization
can be recovered with $M=3\,M_\odot$, $D=10$\,kpc, and $i=75^\circ$.
This is what we essentially now do by considering model fits with the
{\tt kerrbb} model.  The {\tt kerrbb} model \citep{li:05a} uses a
relativistic model of the disk structure, and includes Doppler beaming
and gravitational redshift in calculation of the emergent spectrum.
The spin of the black hole becomes a fit parameter.  Increasing the
spin from 0 (i.e., a Schwarzschild black hole) has two major effects:
it increases the accretion efficiency of the disk, and decreases the
disk inner radius.  These two effects in combination tend to increase
the disk temperature while decreasing its normalization (i.e.,
emitting area or, equivalently, inner radius).  Both of these effects
are suggested in the \fu spectra.

\begin{deluxetable*}{lccccccccc}
\setlength{\tabcolsep}{0.03in} \tabletypesize{\footnotesize}
\tablewidth{0pt} \tablecaption{{\tt
    constant*TBnew*simpl$\otimes$(kerrbb+gaussian)} Fits to \fu
  Spectra \label{tab:kfits}} \tablehead{ \colhead{ObsID} &
  \colhead{$N_\mathrm{H}$} & \colhead{$a^*$} & \colhead{$\dot M$} &
  \colhead{$\Gamma$} & \colhead{$f_\mathrm{scat}$} &
  \colhead{$N_\gamma$} & \colhead{$C_\mathrm{xis0}$} &
  \colhead{$C_\mathrm{xis3}$} & \colhead{$\chi^2/$DoF} \\ &
  ($10^{21}~\mathrm{cm^{-2}}$) & $(Jc/GM^2)$ &
  ($10^{17}~\mathrm{g~s^{-1}}$) & & & $(10^{-4})$ } 
\startdata
\noalign{\vspace*{0.2mm}} 
 405057010 
 & \errtwo{0.185}{0.002}{0.001}    
 & \errtwo{0.8965}{0.0011}{0.0012} 
 & \errtwo{0.180}{0.001}{0.001}    
 & \errtwo{1.1}{2.9}{0.0}          
 & \errtwo{0.000}{0.001}{0.000}    
 & \errtwo{0.9}{0.4}{0.4}          
 & \errtwo{1.019}{0.002}{0.002}    
 & \errtwo{1.020}{0.002}{0.002}    
 & 2062.5/1014 \\ 
\noalign{\vspace*{0.2mm}} 
 405057020 & \nodata & \nodata 
 & \errtwo{0.131}{0.001}{0.001}    
 & \errtwo{1.1}{0.1}{0.0}          
 & \errtwo{0.037}{0.004}{0.004}    
 & \errtwo{0.0}{0.1}{0.0}          
 & \errtwo{1.078}{0.003}{0.003}    
 & \errtwo{1.083}{0.003}{0.003}    
 & \nodata \\ 
\noalign{\vspace*{0.2mm}} 
 405057030 & \nodata & \nodata 
 & \errtwo{0.115}{0.002}{0.001}    
 & \errtwo{4.0}{0.0}{1.3}          
 & \errtwo{0.015}{0.006}{0.008}    
 & \errtwo{0.0}{0.2}{0.0}          
 & \errtwo{1.067}{0.005}{0.004}    
 & \errtwo{1.057}{0.004}{0.005}    
 & \nodata \\ 
\noalign{\vspace*{3mm}} 
 405057010 
 & \errtwo{0.186}{0.002}{0.001}    
 & \errtwo{0.8957}{0.0016}{0.0016} 
 & \errtwo{0.181}{0.001}{0.002}    
 & \errtwo{1.1}{2.9}{0.0}          
 & \errtwo{0.000}{0.002}{0.000}    
 & \errtwo{1.0}{0.5}{0.5}          
 & \errtwo{1.021}{0.003}{0.004}    
 & \errtwo{1.021}{0.003}{0.004}    
 & 1189.8/1014 \\ 
\noalign{\vspace*{0.2mm}} 
 405057020 & \nodata & \nodata 
 & \errtwo{0.131}{0.001}{0.002}    
 & \errtwo{1.1}{0.1}{0.0}          
 & \errtwo{0.032}{0.005}{0.006}    
 & \errtwo{0.0}{0.1}{0.0}          
 & \errtwo{1.077}{0.004}{0.004}    
 & \errtwo{1.081}{0.004}{0.004}    
 & \nodata \\ 
\noalign{\vspace*{0.2mm}} 
 405057030 & \nodata & \nodata 
 & \errtwo{0.116}{0.002}{0.001}    
 & \errtwo{4.0}{0.0}{2.5}          
 & \errtwo{0.009}{0.008}{0.005}    
 & \errtwo{0.0}{0.3}{0.0}          
 & \errtwo{1.067}{0.006}{0.006}    
 & \errtwo{1.056}{0.006}{0.006}    
 & \nodata \\ 
\noalign{\vspace*{2mm}} 
\hline \\ 
 405057010 
 & \errtwo{0.171}{0.002}{0.002}    
 & \errtwo{0.9991}{0.0002}{0.0001} 
 & \errtwo{0.288}{0.007}{0.005}    
 & \errtwo{1.4}{0.1}{0.1}          
 & \errtwo{0.072}{0.009}{0.007}    
 & \errtwo{2.3}{0.5}{0.5}          
 & \errtwo{1.021}{0.003}{0.002}    
 & \errtwo{1.021}{0.003}{0.002}    
 & 1830.0/1014 \\ 
\noalign{\vspace*{0.2mm}} 
 405057020 & \nodata & \nodata 
 & \errtwo{0.224}{0.005}{0.004}    
 & \errtwo{1.1}{0.1}{0.0}          
 & \errtwo{0.150}{0.007}{0.007}    
 & \errtwo{0.1}{0.4}{0.1}          
 & \errtwo{1.080}{0.003}{0.003}    
 & \errtwo{1.085}{0.003}{0.003}    
 & \nodata \\ 
\noalign{\vspace*{0.2mm}} 
 405057030 & \nodata & \nodata 
 & \errtwo{0.176}{0.003}{0.004}    
 & \errtwo{3.4}{0.6}{0.4}          
 & \errtwo{0.127}{0.032}{0.022}    
 & \errtwo{0.7}{0.6}{0.5}          
 & \errtwo{1.069}{0.005}{0.004}    
 & \errtwo{1.058}{0.004}{0.005}    
 & \nodata \\ 
\noalign{\vspace*{3mm}} 
 405057010 
 & \errtwo{0.171}{0.002}{0.002}    
 & \errtwo{0.9993}{0.0002}{0.0002} 
 & \errtwo{0.282}{0.009}{0.007}    
 & \errtwo{1.4}{0.2}{0.1}          
 & \errtwo{0.068}{0.010}{0.008}    
 & \errtwo{2.1}{0.6}{0.5}          
 & \errtwo{1.022}{0.004}{0.003}    
 & \errtwo{1.022}{0.003}{0.004}    
 & 1084.3/1014 \\ 
\noalign{\vspace*{0.2mm}} 
 405057020  & \nodata & \nodata 
 & \errtwo{0.219}{0.007}{0.006}    
 & \errtwo{1.1}{0.1}{0.0}          
 & \errtwo{0.142}{0.009}{0.010}    
 & \errtwo{0.2}{0.5}{0.2}          
 & \errtwo{1.080}{0.004}{0.003}    
 & \errtwo{1.083}{0.004}{0.004}    
 & \nodata \\ 
\noalign{\vspace*{0.2mm}} 
 405057030 & \nodata & \nodata 
 & \errtwo{0.173}{0.004}{0.004}    
 & \errtwo{3.7}{0.7}{0.3}          
 & \errtwo{0.128}{0.026}{0.030}    
 & \errtwo{0.9}{0.6}{0.6}          
 & \errtwo{1.068}{0.007}{0.005}    
 & \errtwo{1.057}{0.006}{0.006}    
 & \nodata 
\enddata

\tablecomments{Errors are 90\% confidence level for one interesting
  parameter.  $N_\mathrm{H}$ is the neutral column; $a^*$ is the black
  hole dimensionless spin; $\dot M$ is the disk accretion rate; {
    $f_{\rm c}$, the disk color-correction factor, has been fixed to
    1.7}; $\Gamma$ and $f_\mathrm{scat}$ are the {\tt simpl}
  parameters (powerlaw slope, constrained to lie between 1.1--4, and
  fraction of scattered continuum); $N_{\gamma}$ is the Gaussian line
  normalization in units of
  photons\,$\mathrm{cm^{-2}}~\mathrm{s}^{-1}$; and $C_\mathrm{xisN}$
  are the normalization constants for the \xis0 and 3 detectors,
  relative to the \xis1 detector.  The top set of parameters are for a
  black hole mass of $3\,M_\odot$ and a distance of 10\,kpc, while the
  bottom set are for a mass and distance of $16\,M_\odot$ and
  22\,kpc. { Within each set of parameters, the first group are for
    data that include only statistical errors, while the second group
    are for data that also include 1.35\% systematic errors.}}
\end{deluxetable*}  

Here we adopt the exact same set of parameters as we considered for
our prior studies of the \rxte spectra of \fu \citep{nowak:08a}.  We
assume a disk inclination of $i=75^\circ$, the theoretically preferred
color-correction value of $f_\mathrm{c}=1.7$ \citep{davis:05a,davis:06a} for
the disk spectrum, and choose two sets of parameters: $3\,M_\odot$ and
10\,kpc, and $16\,M_\odot$ and 22\,kpc.  We discuss these latter two
choice more fully in Section~\ref{sec:scale}.  We further modify these
fits by allowing for a spectral hardening which we model with {\tt
  simpl} \citep{steiner:09a}, which is a convolution model which
mimics some aspects of \emph{optically thin}
Comptonization\footnote{{\tt simpl}, however, should not be confused
  with a Comptonization model.  Aside from lacking any physical
  parameters, e.g., a coronal optical depth, a coronal temperature or
  compactness, and any reference to coronal geometry, it also cannot
  mimic spectral shapes associated with optically thick
  Comptonization.  It should be regarded as purely a phenomenological
  model, albeit one that is more useful than a powerlaw since it does
  not extend its spectrum to low energies in the same manner as a
  steep powerlaw can.}.  The two parameters of this convolution model
are the slope of the powerlaw into which disk photons are transferred,
and the fraction of input photons transferred into the hard tail.

\begin{figure}
\epsscale{1} \plotone{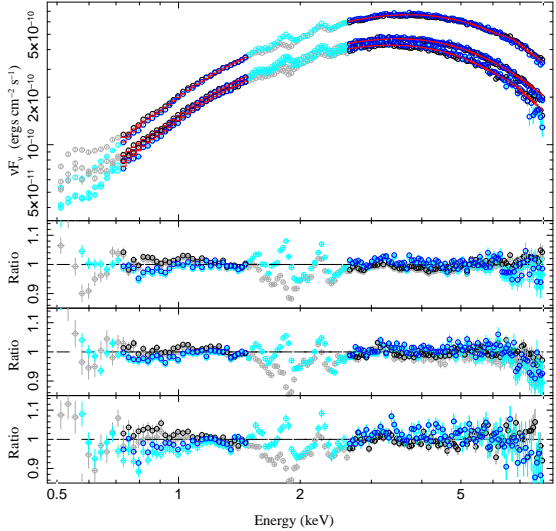}
\caption{The joint fit of a {\tt simpl$\otimes$kerrbb} model to the
  three spectra { assuming a black hole mass and distance of
    $16\,M_\odot$ and 22\,kpc}.  The spectra share a common
  interstellar absorption column, black hole mass and spin, disk
  inclination, { and a fixed color correction factor of $f_{\rm
      c}=1.7$}. They have independent disk accretion rates and {\tt
    simpl} parameters.  Black/grey spectra are the combined \xis0 and
  3 spectra, and blue/light blue are the \xis1 spectra.  Lightly
  shaded data were not included in the fit.  The top residuals panel
  shows the ratio residuals for obsID 405057010, the middle residuals
  panel shows the ratio residuals for obsID 405057020, and the bottom
  shows the ratio residuals for obsID 405057030.}
 \label{fig:kerrbb}
\end{figure}

As for the Comptonization models, we fit all three observations
simultaneously, include a broad Gaussian line in exactly the same
manner, and tie some of the model parameters together.  Specifically,
we constrain each observation to be fit with the same neutral column,
line energy and width (fixed to 0.3\,keV), and black hole spin, but
let the disk accretion rate, the {\tt simpl} model parameters, and the
line normalization be free.  Results for these {\tt kerrbb} model
fits are presented in Table~\ref{tab:kfits} and Fig.~\ref{fig:kerrbb}.
The resulting fits are good, with the $3\,M_\odot$, 10\,kpc model
yielding $\chi^2_\nu/\nu = 2063/1014$ and the $16\,M_\odot$, 22\,kpc
model yielding $\chi^2_\nu/\nu = 1830/1014$.  The latter model is very
comparable to our best Comptonization fit.  Additionally, this latter,
more successful, model has line parameters very similar to those from
the Comptonization fits. We also find that only $\approx 7-15\%$ of
the disk photons are transferred into a hard tail.  The fitted neutral
column is slightly smaller than for the Comptonization fits, but only
by $\lesssim10\%$.

Just as the Comptonization models could fit all three observations
with a constant normalization (equivalent to a constant inner disk
radius), so too can the relativistic disk models fit all three
observations with a constant spin parameter (also equivalent to a
constant inner disk radius).  The main variable parameter among the
observations here becomes the disk accretion rate.  For both sets of
mass and distance assumptions, the fitted spin parameter is high:
nearly 0.9 for the $3\,M_\odot$, 10\,kpc case, and nearly maximally
spinning in the other.  We discuss the scaling relations among the
disk parameters in the next section.

\section{Scaling Relations}\label{sec:scale}

As shown above, spectral fits to the data are somewhat degenerate.  We
have found a successful set of relativistic disk model fits assuming a
mass of $3\,M_\odot$ and a distance of 10\,kpc, and also when assuming
a mass of $16\,M_\odot$ and a distance of 22\,kpc.  The first set of
parameters was chosen as being the lowest mass/closest distance black
hole that is likely to be consistent with \fu persistently remaining
in a soft state.  The second set of parameters was chosen to be
consistent with our prior \rxte studies \citep{nowak:08a}, which in
turn were found via the scaling relations described below.

As the spectra are dominated by the seed photons from a disk for both
the disk and Comptonization models, the flux, $F$, scales as:
\begin{equation}
F \propto \frac{R^2}{D^2}T^4_\mathrm{eff} \propto \frac{M^2}{D^2} \left (
\frac { T_\mathrm{col} }{f_\mathrm{c}} \right )^4 ~~,
\end{equation}
where $D$ is the distance to the source, $R$ is the inner radius of
the disk, $M$ is the black hole mass, $T_\mathrm{eff}$ is the disk's
peak effective temperature, $T_\mathrm{col}$ is the disk's peak color
temperature, and $f_\mathrm{c}$ is the color correction factor.  The
effective temperature to the fourth power scales as the fractional
Eddington luminosity, $L/\mathrm{L_{Edd}}$, multiplied by $M/R^2 \propto
M^{-1}$.  Thus, if we make the assumption that the lowest flux \rxte
observations described by \citet{nowak:08a} are indeed at a fixed
fraction of Eddington luminosity, e.g., 3\%, then we will have:
\begin{equation}
\left ( \frac { T_\mathrm{col} }{f_\mathrm{c}} \right )^4 \propto M^{-1} ~~.
\end{equation}
The color temperature is fixed by observation, and $L/\mathrm{L_{Edd}}
\propto \dot M/M$ (where $\dot M$ is the accretion rate) is fixed by
assumption, leading to the first two of our scaling relations:
\begin{equation}
M \propto f_\mathrm{c}^4 ~~,~~ \dot M \propto f_\mathrm{c}^4 ~~.
\end{equation}

Just as the color temperature is fixed by observations, so too is the
flux.  Given that the disk inner radius scales as mass, and using the
fact that the color temperature and flux are fixed by observation, we
have that $M^2/D^2f_\mathrm{c}^4$ is fixed.  Using the above scaling
relations, we then obtain:
\begin{equation}
D \propto f_\mathrm{c}^2 ~~.
\end{equation}

We have previously used these scaling relations to find spectral fits
for a black hole mass of 16\,$M_\odot$ and a distance of $\approx
22$\,kpc \citep{nowak:08a}.  Specifically, we first explored fits to
\rxte spectra where we fixed the mass and distance to 3\,$M_\odot$ and
10\,kpc. We found that the best fits clustered around a spectral
hardening factor of $f_\mathrm{c} \approx 1.1$.  We then fixed the
spectral hardening factor to $f_\mathrm{c} = 1.7$ and the mass to
16\,$M_\odot$.  We allowed the distance to be a free parameter, and
found that its best fit values clustered around 22\,kpc, i.e., roughly
consistent with the scaling relations above.

\begin{figure}
\epsscale{1} \plotone{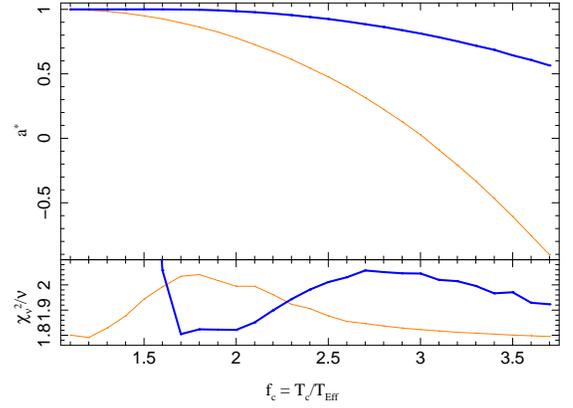}
\caption{Results of fitting a {\tt simpl$\otimes$kerrbb} to the three
  \suzaku observations of \fu.  The top panel shows the fitted spin
  vs. spectral hardening factor.  The bottom panel shows the reduced
  $\chi^2$ value of the fits. The blue { (thicker)} lines represent
  models with the mass and distance fixed to 16\,$M_\odot$ and
  22\,kpc, respectively, and { orange} lines represent models with
  the mass and distance fixed to 3\,$M_\odot$ and 10\,kpc,
  respectively.}
 \label{fig:spin}
\end{figure}

Here we perform a similar exercise for the \suzaku spectra.  As in
Section~\ref{sec:fits}, we fit all three spectra simultaneously with
the {\tt simpl$\otimes$kerrbb} model wherein we only allow the
accretion rate and {\tt simpl} parameters to vary among the three
spectra. We consider two sets of fits, one with mass of 3\,$M_\odot$
and distance 10\,kpc, and another with 16\,$M_\odot$ and distance
22\,kpc.  For both we allow the black hole spin to be a free
parameter, and we create grids of fits for fixed spectral hardening
factors ranging over $f_\mathrm{c}=1.1$--3.7.  The plot of fitted
spin, $a^*$, vs. spectral hardening factor, $f_\mathrm{c}$, is
presented in Fig.~\ref{fig:spin}.

As for our prior \rxte observations, we find that fits using
$3\,M_\odot$, 10\,kpc, and $f_\mathrm{c}=1.1$ or $16\,M_\odot$,
22\,kpc, and $f_\mathrm{c}=1.7$ are nearly equally good and are of
comparable $\chi^2$ to our best Comptonization fits.  We also see that
all fits using $16\,M_\odot$, and 22\,kpc require large spin
parameters.  The fits using $3\,M_\odot$ and 10\,kpc only achieve low
spin parameters for large color correction factors,
$f_\mathrm{c}\gtrsim3$.  { (These conclusions are unaltered if
  uniform 1.35\% systematic errors are included in the fits.)}

\section{Discussion}\label{sec:discuss}

There are four salient features of these \suzaku observations of \fu
that echo the results that we obtained with our prior \chandra, \xmm,
and \rxte studies.  The peak disk temperatures are high, the disk
normalizations (or equivalently, emitting areas or inner disk radii)
are small, all observations can be fit with the same inner disk
radius, and the spectra are remarkably simple.  The only significant
deviations from a simple continuum, aside from spectral hardening and
a possible weak, broad Fe line at high energy, are
$\approx0.7$--$0.9$\,keV features associated with absorption by the
ISM, predominantly due to the Fe L and Ne edges.

\begin{figure}
\epsscale{1} \plotone{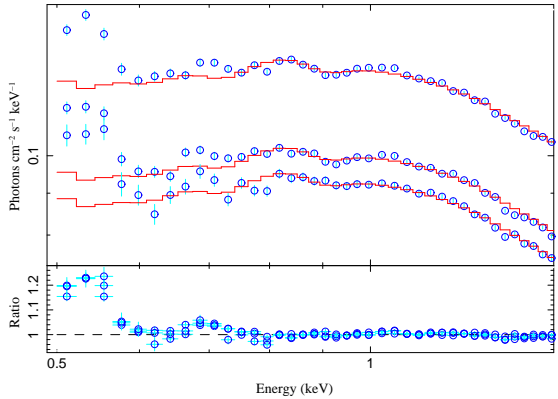}
\caption{The same fit as seen in Fig.~\protect{\ref{fig:eqpair}}, but
  showing the \xis0, 1, and 3 spectra combined, and the model applied
  down to 0.5\,keV.}
 \label{fig:edge}
\end{figure}

We show a closeup of this region in Fig.~\ref{fig:edge}.  The spectral
feature seen at $\approx 0.55$\,keV, not described by our fitted
models, can in fact be seen in the flux corrected spectra of
\emph{any} monochromatic, narrow line spectrum simulated using the
\suzaku response functions. Small calibration errors in the amplitude
of this feature potentially could lead to a systematic underestimate
or overestimate of the predicted counts near 0.55\,keV.  We thus do
not believe that this feature is indicative of physical complexity in
the underlying \fu spectra.  Likewise, it is unclear whether the
$\approx 5\%$ residuals seen near the 0.7\,keV Fe L edge represent
complexity in the physical spectrum, as opposed to uncertainty in the
response. Aside from the Fe L edges, \chandra and \xmm spectra did not
show any spectral features near the 0.7\,keV bandpass
\citep{nowak:08a}.

The only clear spectral features are the Ne edge/Ne {\sc ix}
absorption near 0.9\,keV.  Similarly simple disk spectra were implied
by \suzaku observations of LMC~X-3 in its soft, disk dominated state
\citep{kubota:10a}.  As pointed out by \citet{kubota:10a}, such
spectral simplicity is in fact surprising if modern disk atmosphere
models apply to these sources.  These disk atmosphere models predict
spectral structure with $\pm5\%$ amplitude near metal edges
\citep{davis:06b} that is seen neither here nor in \suzaku observations of
LMC~X-3.  That the simplest, phenomenological disk models lacking such
features describe the spectra extremely well (i.e., to within
$\lesssim5$\% 0.7--0.8\,keV, and to $\lesssim2\%$ above 0.8\,keV) is
quite remarkable, and may indicate that additional physics needs to be
incorporated into the disk atmosphere models \citep{kubota:10a}.

We also have confirmed the results of our prior studies in that the
best fits assuming a small mass and distance (3\,$M_\odot$, 10\,kpc)
require a black hole spin $a^* \gtrsim 0.9$.  The required spin
increases if we instead assume 16\,$M_\odot$ and 22\,kpc.  Altering
the assumed disk inclination does not change this situation.  Higher
disk inclinations are ruled out by the lack of system eclipses.  Lower
disk inclinations allow for more observed gravitational redshifting,
{ and less relativistic beaming,} of the spectrum, and thus require
larger spectral correction factors.  {(I.e., inclination effects
  more than just the projected area of the disk.)} As an example, we
took the 16\,$M_\odot$ and 22\,kpc \texttt{kerrbb} model, froze the
inclination to $45^\circ$ and let both the black hole spin and
hardness correction factor become free parameters.  A solution with
$\chi^2$ comparable to the best fits discussed above was found,
however, it required \emph{maximally negative spin} of $a^* \approx
-1$ and an extremely large color correction factor of $\approx 8$.
Thus, high spin/high inclination solutions are the most
``straightforward'' solutions that we have been able to obtain, {
  and generally have had the lowest color-correction factors of the
  models that we have explored \citep[see also][]{nowak:08a}.}

These results are in fact confirmed by the Comptonization model fits:
the high disk temperatures (1.31--1.48\,keV) and especially the low
normalization --- 15 times lower than expected for a $3\,M_\odot$
Schwarzschild black hole at 10\,kpc --- could be indicative of a
rapidly spinning black hole.  As discussed above, the disk temperature
in the {\tt diskpn} seed photons used in the Comptonization model
peaks at $\approx 10\,GM/c^2$.  If one were to reduce this radius to
$\approx 2.5\,GM/c^2$, i.e., change the emission profile to more
closely match that of a rapidly spinning hole, then the Comptonization
model normalization would be consistent with a $3\,M_\odot$ black hole
at only 10\,kpc.

The Comptonization models offer two possibilities\footnote{Note that
  we are ignoring issues of the unknown inclination.  Choosing a more
  face-on inclination only exacerbates the normalization problem
  (i.e., we would have expected an even larger normalization), and
  would introduce more gravitational redshifting to the observed
  temperature, increasing the requirement for rapid spin in the
  spectral fits.} for fitting the \fu spectrum with a low spin. The
first, perhaps less likely, possibility is to increase the
color-correction factor via a low temperature, optically thick corona.
We have already seen that the lowest flux observations are fit with
such coronae, albeit with a Compton $y$-parameter of $\approx 0.1$.
Instead, we are envisioning a scenario wherein $y \approx f_\mathrm{c}
\approx 3$ -- the color correction value required to fit a
$3\,M_\odot$, 10\,kpc, Schwarzschild black hole.  This would require
an optical depth of
\begin{equation}
\tau \approx 9.8 \left ( \frac{4\,\mathrm{keV}}{kT_\mathrm{c} }\right )^{1/2} \left (
\frac{ f_\mathrm{c} } {3} \right )^{1/2} ~~,
\end{equation}
where $kT_\mathrm{c}$ is the coronal temperature.  This is a coronal
regime typically not explored in spectral models.  The concept would
be that the low flux, equilibrium solution of the disk was such a
cool, optically thick atmosphere which at higher flux (i.e., the 15\%
of the time requiring a hard tail) expanded into a hot, optically thin
corona.

Aside from lacking a theoretical description of such a scenario, the
fitted constant emitting area at low flux implies that the \fu system
must always return to the \emph{same} equilibrium coronal solution at
low flux.  It is not clear why this should be the case.  The second
possibility offered by the Comptonization solutions is simply to
increase the distance to \fu, while keeping the mass small.  The
scaling relations discussed in Section~\ref{sec:scale} only apply
under the assumption that the lowest flux observations are a fixed
fraction of the Eddington luminosity (e.g., 2\%\,$\mathrm{L_{Edd}}$).
If we relax this assumption, then eq.~\ref{eq:scale} suggests that if
the \fu distance were 40\,kpc, then it could be a $3\,M_\odot$
Schwarzschild black hole.  This would imply, however, that the
\emph{faintest} \fu spectra are $\gtrsim 30\%\,\mathrm{L_{Edd}}$,
while the brightest observed \rxte spectra are $\gtrsim
\mathrm{L_{Edd}}$.  This would be rather unusual, especially given the
lack of strong variability found in the \rxte spectra: the root mean
square variability is typically $\lesssim 3\%$, and at its strongest
was $\lesssim 13\%$ in the 2--60\,keV bandpass over
$\approx10^{-2}$--$2\times10^3$\,Hz \citep{nowak:99d,nowak:08a}.

It therefore seems likely that 4U 1957+11 is a system that has a
combination of both high spin and large distance, which puts this
X-ray binary in the Galaxy's halo. It is interesting to speculate
whether these two characteristics are related to one another. First,
we note the obvious fact that in order for 4U 1957+11 to be a X-ray
binary it must still be accreting and therefore be a comparatively
young system. Given the low star formation rate in the Galaxy's halo,
it is rather unlikely, therefore, that 4U 1957+11 was formed in the
halo. Instead 4U 1957+11 likely originates in the Galactic disk and
then migrated into the halo. Several possible formation scenarios have
been discussed in the literature, most recently associated with the
discovery of young hypervelocity stars in the Galaxy's halo. For
example, \citet[][and references therein]{perets:09a} discuss a
scenario where hypervelocity binary systems are created through the
interaction of a hierarchical triple star system with the supermassive
black hole in the center of the Galaxy. Since hierarchical triples are
very common among multiple star systems, such a scenario is likely.
With a typical ejection velocity of
1800\,$\mathrm{km}\,\mathrm{s}^{-1}$, it would take this system a few
$10^6$\,years to reach its current position, well within the lifetime
of typical evolution scenarios for X-ray binary systems
\citep{vandenheuvel:76a}.  { (At such a velocity, and assuming a
  22\,kpc distance, the proper motion of \fu likely would have
  amounted to $\aproxlt 0.3^{\prime\prime}$ since the discovery of the
  optical counterpart by \citealt{margon:78a}.)}  Such a scenario
could also explain a large spin for the system, if during the ejection
angular momentum is transferred onto the black hole.

A high black hole spin for \fu is a consistent interpretation of the
results presented here; however, such a result is dependent upon the
unknown mass, distance, and inclination of the system. Given the
nature of \fu as perhaps the simplest, cleanest example of a BHC soft
state, and perhaps the most rapidly spinning black hole, observations
to independently determine this system's parameters are urgently
needed.

\acknowledgements Michael Nowak was supported by NASA Grants
NNX10AR94G and SV3-73016. J\"orn Wilms was partly supported by the
European Commission under contract ITN 215212 ``Black Hole Universe''
and by the Bundesministerium f\"ur Wirtschaft und Technologie through
Deutsches Zentrum f\"ur Luft- und Raumfahrt grants 50OR0701 and
50OR1005.


\end{document}